\documentclass[usenatbib]{mnras}

\usepackage{amsmath, amssymb, tikz}
\usepackage{color}

\title[The absence of a thin accretion disc in M81]{The absence of a thin disc in M81}

\author[A. J. Young, et al.]{A. J. Young$^1$\thanks{E-mail:
Andy.Young@bristol.ac.uk}, I. McHardy$^2$, D. Emmanoulopoulos$^2$ and S. Connolly$^2$\\
$^{1}$H.H. Wills Physics Laboratory, Tyndall Avenue, Bristol BS8 1TL.\\
$^{2}$Department of Physics and Astronomy, University of Southampton, University Road, Southampton SO17 1BJ.}

\begin{document}

\pagerange{\pageref{firstpage}--\pageref{lastpage}} \pubyear{2016}

\maketitle

\label{firstpage}

\begin{abstract}
We present the results of simultaneous \emph{Suzaku} and \emph{NuSTAR} observations of the nearest Low-Luminosity Active Galactic Nucleus (LLAGN), M81$^{*}$. The spectrum is well described by a cut-off power law plus narrow emission lines from Fe~K$\alpha$, \ion{Fe}{xxv} and \ion{Fe}{xxvi}. There is no evidence of Compton reflection from an optically thick disc, and we obtain the strongest constraint on the reflection fraction in M81$^*$ to date, with a best-fit value of $R = 0.0$ with an upper limit of $R < 0.1$. The Fe~K$\alpha$ line may be produced in optically thin, $N_H = 1 \times 10^{23} \text{\,cm}^{-2}$, gas located in the equatorial plane that could be the broad line region. The ionized iron lines may originate in the hot, inner accretion flow. The X-ray continuum shows significant variability on $\sim 40$\,ks timescales suggesting that the primary X-ray source is $\sim 100$s of gravitational radii in size. If this X-ray source illuminates any putative optically thick disc, the weakness of reflection implies that such a disc lies outside a $\text{few} \times 10^3$ gravitational radii. An optically thin accretion flow inside a truncated optically thick disc appears to be a common feature of LLAGN that are accreting at only a tiny fraction of the Eddington limit.
\end{abstract}


\section{Introduction}

M81, also known as NGC 3031, is a spiral galaxy hosting the closest Low-Luminosity Active Galactic Nucleus (LLAGN), M81$^{*}$, at a distance of 3.63~Mpc \citep{Freedman1994}. Its central black hole has a mass of $M = 7 \times 10^7 M_\odot$ \citep{Devereux2003}. The bolometric luminosity of M81$^*$ is $9.3 \times 10^{40} \text{\,erg\,s}^{-1}$ \citep{Ho1996}, which corresponds to $1.1 \times 10^{-5} L_\text{Edd}$, where $L_\text{Edd}$ is the Eddington luminosity. Thus, logarithmically speaking, M81$^{*}$ is midway between ultra-low Eddington fraction accreters such as Sgr~A* at the centre of the Galaxy with $L_\text{Sgr\ A*} \sim 3 \times 10^{-10} L_\text{Edd}$ \citep{Melia2001} and a typical high-efficiency AGN with $L_\text{AGN} \sim 0.1 - 1 L_\text{Edd}$. M81$^*$ is particularly interesting to study because it is nearby and bright enough to allow detailed spectroscopy, while being in the sub-Eddington regime in which the accretion flow is thought to be qualitatively different to that of an efficiently radiating, high accretion rate AGN.

Orbiting the black hole in M81$^*$ at large, $\sim 10$~pc, scales is a disc inclined at $14^\circ$ to the line of sight (i.e., the angle between the normal to the disc and the observer is $14^\circ$) as measured from its narrow optical emission lines \citep{Devereux2003}. On smaller scales, there is a 0.8~pc radius ($2.5 \times 10^5 r_g$, where $r_g = GM/c^2$) disc inclined at $50^\circ$ to the line of sight, the presence of which is inferred from its broad H$\alpha$ line \citep{Devereux2007}. The inclination angle of this inner disc is consistent with that of the short, 0.017~pc long, radio jet from the nucleus which is also inclined at $50^\circ$ \citep[i.e., the jet is perpendicular to the inner disc;][]{Bietenholz2000}. These observations suggest that there is a warp between the pc and 10~pc-scale discs.

The thin disc does not extend all the way into the event horizon of the black hole. While X-ray observations with \emph{ASCA} \citep{Ishisaki1996} revealed a possibly broadened Fe~K line with an Equivalent Width (EW) of 170~eV, no strong Compton reflection component was detected. Subsequent observations with \emph{BeppoSAX} detected a broad complex of Fe lines with the same EW, 170~eV, but again no Compton reflection component. The \emph{BeppoSAX} spectrum extended to over 100~keV allowing the reflection fraction $R = \Omega / 2\pi$,  where $\Omega$ is the solid angle subtended by the optically thick gas from the point of view of the X-ray continuum source, to be constrained to be $R < 0.3$ \citep{Pellegrini2000}. More recent observations with \emph{XMM-Newton} were able to resolve the broad Fe~K complex into three separate lines of almost equal strength due to Fe~K$\alpha$, \ion{Fe}{xxv} and \ion{Fe}{xxvi} with EWs of 39, 47 and 37 eV, respectively \citep{Page2004}. High spectral resolution observations with the \emph{Chandra} High Energy Transmission Gratings (HETGS) spectrometer showed that the \ion{Fe}{xxvi} line is redshifted by $\sim 2500 \text{\,km\,s}^{-1}$ while the neutral Fe K$\alpha$ line, with an EW of 47~eV, is consistent with being at rest \citep{Young2007}. The Fe K$\alpha$ line is unresolved, with a Full Width at Half Maximum (FWHM) of $< 4670 \text{\,km\,s}^{-1}$. For a disc inclined at $50^\circ$ this requires $r_\text{in} \gtrsim 10^4 r_g$. For a face-on disc, however, the constraint is much weaker and formally fitting a {\tt diskline} model \citep{Fabian1989} only constrains $r_\text{in} > 55 r_g$ \citep{Young2007}. The Si~K$\alpha$ line appears to be broadened, with a FWHM $\sim 1433 \text{\,km\,s}^{-1}$ which corresponds to $r_\text{in} \simeq 10^5 r_g$ in a disc inclined at $50^\circ$. This radius is comparable to that of the broad H$\alpha$ emitting gas. In this truncated disc model the inner accretion flow is optically thin, radiatively inefficient and may produce the ionized \ion{Fe}{xxv} and \ion{Fe}{xxvi} emission lines, characteristic of emission from plasma at temperatures of $10^7 - 10^8 \text{\,K}$ close to the truncation radius \citep[see][for more details of this model]{Young2007}. The origin of the Fe~K$\alpha$ line is less straightforward.

X-ray monitoring of M81$^*$ has shown the continuum power-law photon index to be anti-correlated with X-ray flux, so the source is harder when it is brighter \citep{Miller2010, Connolly2016}. This is consistent with models in which the continuum is produced by the inner accretion flow which Comptonises seed photons from the inner edge of truncated disc, with changes in photon index associated with changes of the truncation radius \citep[e.g.][]{SKipper2013}.

In this paper we address the questions of where the Fe~K$\alpha$ line originates, what is the geometry of the inner accretion flow, and what is the transition radius between the outer thin disc and inner thick disc. To answer these questions we make use of simultaneous \emph{Suzaku} and \emph{NuSTAR} observations to measure the strength of the Fe~K$\alpha$ line, Compton reflection hump and hence reflection fraction. In this way we hope to disentangle whether the Fe~K$\alpha$ line is produced by scattering in optically thin gas, or is produced by reflection from an optically thick disc.

M81 has a redshift of $z = -0.0001$. The Galactic column density in this direction is $N_H (\text{Gal}) = 4.22 \times 10^{20} \text{\,cm}^{-2}$.

\section{Observations}

\begin{figure}
    \centering
    \includegraphics{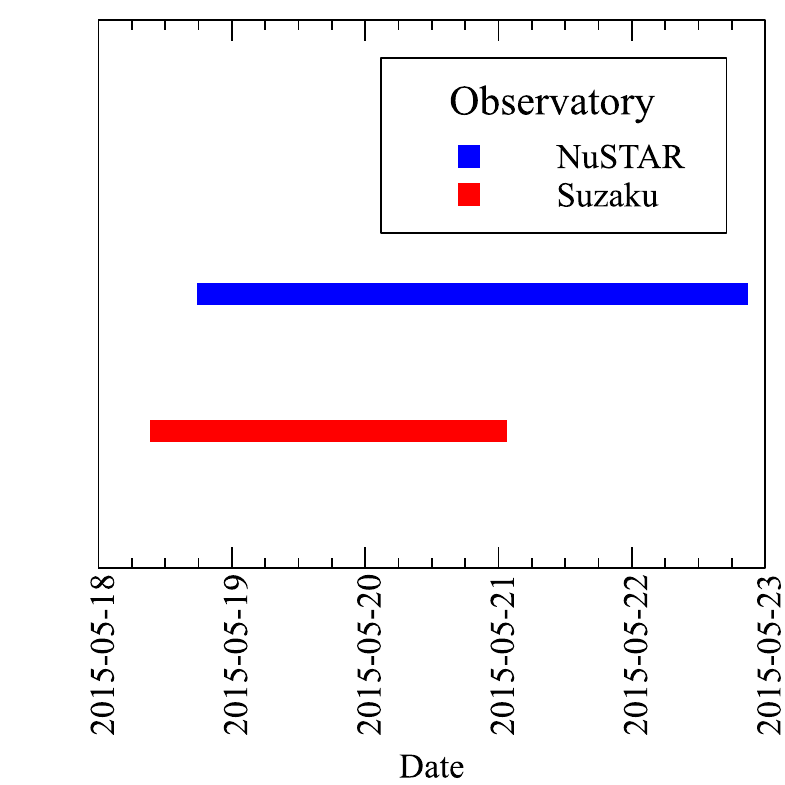}
    \caption{Date and time of observations of M81 with \emph{NuSTAR} and \emph{Suzaku}. There are about two days of strictly simultaneous observation with both observatories.}
    \label{fig:observations}
\end{figure}

The X-ray data presented here come from a coordinated \emph{NuSTAR} and \emph{Suzaku} observing programme in May 2015. The observing times are simultaneous for about two days, while the \emph{NuSTAR} observations extend for a further two days, as illustrated in Fig.~\ref{fig:observations} and summarised in Table~\ref{tab:observations}. In addition, we have contemporaneous \emph{Swift} X-ray and \emph{AMI} radio observations that will be presented elsewhere.

\begin{table}
    \centering
    \begin{tabular}{ccc}
        \hline
        Observatory & Start date & Total good time \\
        \hline
        \emph{NuSTAR} & 2015-05-18 & 209 ks \\
        \emph{Suzaku} & 2015-05-18 & 99 ks \\
        \hline
    \end{tabular}
    \caption{Observations of M81 taken with \emph{NuSTAR} and \emph{Suzaku} showing the start time and total good time (averaged over the active instruments for each observatory).}
    \label{tab:observations}
\end{table}

The X-ray data were extracted and analysed using HEAsoft version 6.18, using standard processing techniques as described below.

\subsection{NuSTAR}

The \emph{NuSTAR} data, from ObsID 60101049002, were processed using CALDB version 20150702. The standard \emph{NuSTAR} pipeline processing tools, {\tt nupipeline} and {\tt nuproducts}, were used to produce spectra and light curves. The on-source time was 343~ks which resulted in a total good exposure time of 209~ks, where the difference between these two times is mostly due to the lost exposure time caused by Earth occultations and passages through the South Atlantic Anomaly (SAA) each orbit. Source counts were extracted from a circular aperture of radius $100^{\prime\prime}$ centred on M81$^*$, with background counts taken from a source-free circular aperture of the same radius to the north of the nucleus. The spectra from modules A and B were fit jointly, rather than being combined.

\begin{figure}
    \centering
    \includegraphics[width=8cm]{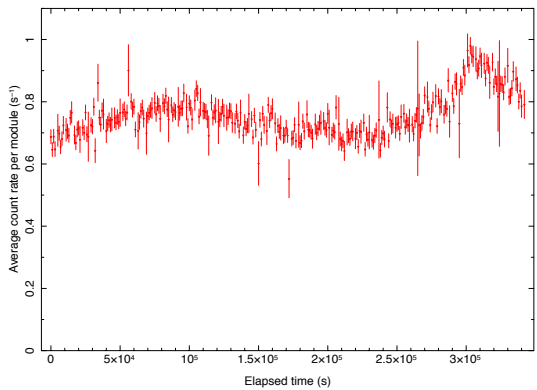}
    \caption{\emph{NuSTAR} full-band, 3--79~keV, background subtracted light curve of M81$^*$ averaging the count rates of modules A and B. Significant, $\sim 30\%$, variability is seen on timescales of $\sim 40 \text{\,ks}$.}
    \label{fig:lc}
\end{figure}

The \emph{NuSTAR} full-band, 3--79~keV, background subtracted light curve in 1000\,sec bins was produced using the pipleline, and is shown in Fig.~\ref{fig:lc}. The error in the source and background counts was given by $\sqrt N$, where $N$ is the number of counts in a bin, and these errors were propagated through the calculation to give the error bars shown in the figure.

\subsection{Suzaku}

The \emph{Suzaku} data, from ObsID 710017010, were processed using the CALDB with XIS calibration products dated 2015-01-05 and XRT mirror calibration products dated 2011-06-30. The HXD was not used during this observation. The on-source time was 217~ks which resulted in a total good exposure time of 98~ks for XIS~0 and 100~ks for XIS 1 and 3. The good time is significantly shorter than the observation duration due to the low Earth orbit of the satellite resulting in Earth occultations and passage through the SAA. Source counts were extracted from a circular aperture of radius $260^{\prime\prime}$ centred on M81$^*$, excluding the region within $100^{\prime\prime}$ of the Ultra-Luminous X-ray Source (ULX) M81 X-6 that is $\sim 200^{\prime\prime}$ south of the nucleus. Background counts were taken from a source-free circular aperture of radius $150^{\prime\prime}$ to the northeast of the nucleus. The spectra from XIS 0, 1 and 3 were not combined, but fit jointly.

Note that during these observations there was a problem with charge leakage in XIS~0 that resulted in telemetry saturation and therefore some data drop-outs that were taken care of with the pipeline processing. This is likely related to the expansion of the dead area of XIS~0 as described in the ISAS memo JX-ISIS-SUZAKU-MEMO-2015-05\footnote{\url{http://www.astro.isas.jaxa.jp/suzaku/doc/suzakumemo/suzaku_memo_2015-05.pdf}.}.

\subsection{Spectral fitting}

We simultaneously fit the three \emph{Suzaku} XIS spectra and the two \emph{NuSTAR} spectra. To account for calibration differences between instruments and detectors the spectral model is multiplied by a (different) constant for each detector. We find that the two \emph{NuSTAR} telescope modules agree very well. If the normalisation of FPMA is fixed at unity, the normalisation of FPMB is $\sim 1.02$, well within the anticipated calibration uncertainty\footnote{See, e.g., the analysis caveats at \url{http://heasarc.gsfc.nasa.gov/docs/nustar/analysis/}.}. With \emph{Suzaku} the XIS~1 and XIS~3 detectors are in good agreement with one another, and within about 4\% of the \emph{NuSTAR} detector normalisation. The constant used for XIS~1 is $\sim 0.94$ and the constant for XIS~ 3 is $\sim 0.98$. The constant for XIS~0, however, is $\sim 0.85$, which is lower than expected but likely linked to problems with the XIS~0 charge leakage. However, there is no apparent difference in the \emph{shape} of the spectrum from XIS~0 when we compare the residuals of fit models to individual detectors. The precise values of these normalisations change slightly from fit to fit. Unless otherwise stated, the data are binned such that each point in the spectrum from an individual satellite has a signal to noise ratio of at least 20; the background is included in this calculation.

\section{Results}

\subsection{Light curve} \label{sec:light_curve}

The full-band, 3--79~keV, background subtracted \emph{NuSTAR} light curve is shown in Fig.~\ref{fig:lc}, and shows that there is significant, $\sim 30\%$, variability on timescales of $\sim 40 \text{\,ks}$. For a black hole mass of $7 \times 10^7 M_\odot$, 40\,ks corresponds to the light crossing time of $116 r_g$. This suggests that a significant fraction of the X-ray continuum originates in a very compact region within a few hundred gravitational radii of the black hole.

\subsection{Simple spectral models}

If we model the 0.5--30~keV continuum as an absorbed power law we find significant residuals in the 6--7~keV band due to iron. Excluding the 5--8~keV band and refitting we obtain a good description of the continuum with $\Gamma = 1.89^{+0.01}_{-0.01}$ and $N_H = 6.9^{+0.2}_{-0.2} \times 10^{20} \text{cm}^{-2}$, having $\chi^2 / \text{d.o.f.} = 2545 / 2193 = 1.16$ (where d.o.f. means degrees of freedom). The error bars on parameter values give the 90\% statistical confidence interval for one interesting parameter (i.e., corresponding to $\Delta \chi^2 = 2.71$).

\begin{figure}
\centering
    \includegraphics[width=8cm]{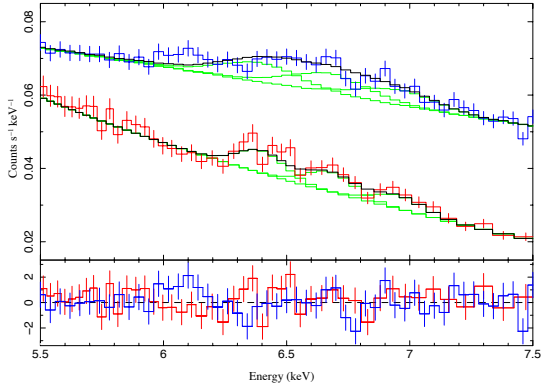}
    \caption{The Fe K band of M81$^*$ from the \emph{NuSTAR} (blue points) and \emph{Suzaku} (red points) observations. The upper panel shows the data, model (black line), and model components folded through the instrument responses (green lines) while the lower panel shows the residuals. Three narrow Gaussian lines have been added at 6.40 keV, 6.68 keV and 6.96 keV representing neutral Fe K$\alpha$, \ion{Fe}{xxv} and \ion{Fe}{xxvi}, respectively. The \emph{Suzaku} data are able to cleanly separate these lines.}
    \label{fig:fe_lines}
\end{figure}

We now include the Fe K band. The iron lines are described by narrow Gaussian lines fixed at 6.40~keV, 6.68~keV and 6.96~keV, representing neutral Fe K$\alpha$, He-like \ion{Fe}{xxv} and H-like \ion{Fe}{xxvi}, respectively. This fit is shown in Fig.~\ref{fig:fe_lines}, and the model parameters for all fits are given in Table~\ref{tab:fits}. The normalisations of each of the lines are free parameters. The spectrum continues to be well described by a power law of photon index $\Gamma = 1.89^{+0.00}_{-0.01}$, column density $N_H = 6.9^{+0.2}_{-0.3} \times 10^{20} \text{cm}^{-2}$,  with line strengths and equivalent widths as given in Table~\ref{tab:fe_lines}. This is a good fit, with $\chi^2 / \text{d.o.f} = 2933 / 2550  = 1.15$. The \emph{Suzaku} data allow us to accurately measure the strength of the neutral Fe K$\alpha$ line while cleanly separating the contribution of \ion{Fe}{xxv} and \ion{Fe}{xxvi} in this band. This is important because the neutral reflection we wish to constrain is only expected to contribute to the neutral Fe K$\alpha$ line. The 0.5--10~keV model flux and luminosity are $3.2 \times 10^{-11} \text{erg cm}^{-2} \text{ s}^{-1}$ and $5.1 \times 10^{40} \text{erg s}^{-1}$, respectively.

\begin{table}
    \centering
    \begin{tabular}{cccc}
        \hline
        Energy & Line & Normalisation & EW \\
        (keV) & ID & ($\times 10^{-6} \text{ ph cm}^{-2} \text{ s}^{-1}$) & (eV) \\
        \hline
        6.40 & Fe K$\alpha$ & $8.5^{+1.6}_{-1.6}$ & $40^{+8}_{-7}$ \\
        6.68 & \ion{Fe}{xxv} & $7.6^{+1.7}_{-1.6}$ & $38^{+8}_{-9}$ \\
        6.96 & \ion{Fe}{xxvi} & $6.0^{+1.5}_{-1.5}$ & $33^{+8}_{-9}$ \\
        \hline
    \end{tabular}
    \caption{Fe K lines in M81* measured from joint fits to the \emph{NuSTAR} and \emph{Suzaku} data. The fourth column shows the Equivalent Width (EW) of each line.}
    \label{tab:fe_lines}
\end{table}

\begin{figure}
    \centering
    \includegraphics[width=8cm]{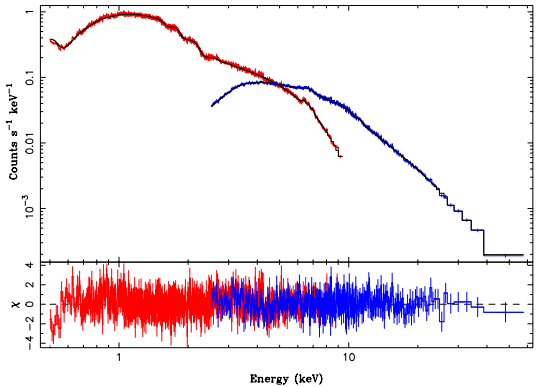}
    \caption{\emph{Suzaku} (red) and \emph{NuSTAR} (blue) spectra of M81$^*$, with the best-fit absorbed, cut-off power law plus three narrow Gaussian iron lines shown as the solid line. Residuals to this fit are in the lower panel. There is no evidence of a Compton reflection hump at high energies.}
    \label{fig:spec}
\end{figure}

While the signal to noise ratio is dominated by the lowest energies, there is no strong evidence of a systematic deviation from the power law model at high energies, other than a slight hint of spectral softening above $\sim 15 \text{\,keV}$. This softening can be modelled by a cut-off power law with an e-folding energy of $267^{+133}_{-67} \text{\,keV}$, with this one additional free parameter improving the fit by $\Delta \chi^2 = 25$, although the reduced $\chi^2$ value is essentially unchanged. There is, however, no clear evidence of any \emph{upturn} in the spectrum at higher energies as one might expect if there were a strong Compton hump. At soft X-ray energies, a small amount of absorption is required in addition to the Galactic column, and this intrinsic absorption has been reported previously \citep[e.g.,][]{Page2003}. The best-fit cut-off power law plus iron line model is shown in Fig.~\ref{fig:spec}.

\subsection{Optically thick reflection modelling of the continuum and iron lines}

\begin{figure}
    \includegraphics[width=0.45\textwidth]{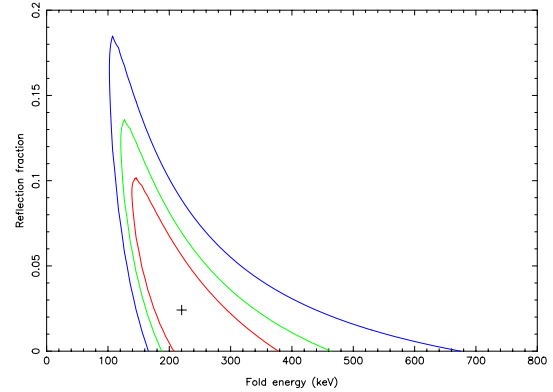}
    \caption{Constraints on the reflection fraction and power-law cut-off energy in M81$^*$. The data are consistent with no reflection. The power-law cut-off energy is constrained to be $E_\text{cut} = 220^{+173}_{-86} \text{\,keV}$. Contours show the 68\%, 90\% and 99\% confidence regions in red, green and blue respectively, with a 90\% confidence upper limit on $R < 0.11$. The best-fit value is shown by the `$+$' symbol.}
    \label{fig:reflection}
\end{figure}

The continuum modelling described above does not suggest that there is any strong Compton reflection hump in the spectrum of M81$^*$. We can measure the strength of any putative reflection continuum from optically thick material by including a Compton reflection component to the spectral model. We replace the cut-off power-law model with the {\sc pexrav} model in which the normalisation, photon index, e-fold energy and reflection fraction are free parameters. The disc inclination is fixed at $50^\circ$, although allowing this to be a free parameter does not change our conclusions. We find that the best fit value for $R$ is consistent with zero, i.e., no reflection, with a 90\% upper confidence limit of $R < 0.11$, where $R = 1$ corresponds to an isotropic point source above an infinite disc covering $2 \pi \text{\, sr}$ of the sky. The power-law cut-off energy is constrained to be $E_\text{cut} = 220^{+173}_{-86} \text{\,keV}$. Contours of the constraints on these two parameters are shown in Fig.~\ref{fig:reflection}. Thus the K$\alpha$ line cannot be produced by reprocessing in optically thick material.

\subsection{Optically thin reprocessing modelling of the continuum and iron lines}

An alternative scenario to reflection from Compton thick material for producing the neutral Fe~K$\alpha$ line is that the line is produced by reprocessing in optically thin material which does not produce a strong Compton hump. We model the continuum, scattered and iron line components from a centrally illuminated torus using the {\sc mytorus} model applied to a power-law continuum with a 200~keV high energy cut-off \citep[see][and the associated instruction manual for a full description of how this model is supposed to be used]{Murphy2009, Yaqoob2012}. The geometry is that of a torus with half-opening angle fixed at $60^\circ$, and we find an acceptable fit for viewing angle $i < 60^\circ$, i.e., the view of the nucleus is unobscured by the torus. Note that acceptable fits might be possible with different opening angles for the torus but the version of {\sc mytorus} we used has a fixed opening angle and an acceptable description could be found with this value. The iron K$\alpha$ line is then produced by scattering of the $\Gamma = 1.9$ power-law continuum by the optically-thin\footnote{We consider $N_H < 10^{24} \text{\,cm}^{-2}$ to be optically thin.} torus with a column density $N_H = 1.1^{+0.3}_{-0.2} \times 10^{23} \text{\,cm}^{-2}$. This provides a good description of the spectrum, with $\chi^2 / \text{d.o.f.} = 2837 / 2529 = 1.12$.

The ``torus'' might correspond to the Broad Line Region (BLR). In M81$^*$, $1 \text{\,pc} = 3 \times 10^5 r_g$, so the size of the torus may be comparable to the size of the BLR which is particularly large in M81 with a radius of maybe $\sim 1 \text{ pc}$ \citep{Devereux2007}. So this hypothesis is compatible with the optical properties of the galaxy. The BLR might very well have a ``torus''-like geometry because both disc and spherical geometries are able to explain the optical lines \citep{Devereux2007}. The column density $N_H \simeq 1 \times 10^{23} \text{ cm}^{-2}$ is a reasonable column densify for a BLR cloud \citep[e.g.,][]{Peterson2006}. The Fe~K$\alpha$ line and velocity resolved Si~K$\alpha$ line \citep[the latter reported by][]{Young2007} could both originate in the BLR. In this model the highly ionised \ion{Fe}{xxv} and \ion{Fe}{xxvi} lines would be associated with the inner hot, inefficient flow within a few thousand $r_g$. \citet{Bianchi2008} reached a similar conclusion for NGC~7213. In some AGN, such as NGC~5548, the BLR can contribute significantly to the X-ray / optical lags \citep[e.g.,][]{Fausnaugh2016}.

\begin{table*}
    \begin{tabular}{cccccccc}
        \hline
        Model & $\Gamma$ & $N_H$ & $i$ & $R$ & $E_\text{cut}$ & Norm & $\chi^2 / \text{d.o.f.}$\\
        & & (cm$^{-2}$) & ($^\circ$) & & (keV) & & \\
        \hline
        Powerlaw & $1.89^{+0.00}_{-0.01}$ & $6.9^{+0.2}_{-0.3} \times 10^{20}$ & & & & $7.1 \pm 0.1 \times 10^{-3}$ & 2933 / 2550 \\
        \hline
        Cut-off powerlaw & $1.86^{+0.01}_{-0.01}$ & $6.3^{+0.3}_{-0.3} \times 10^{20}$ & & & $267^{+133}_{-67}$ & $7.0^{+0.1}_{-0.1} \times 10^{-3}$ & 2908 / 2549 \\
        \hline
        {\tt pexrav} & $1.86^{+0.01}_{-0.01}$ & $6.3^{+0.3}_{-0.3} \times 10^{20}$ & & $< 0.11$ & $220^{+173}_{-86}$ & $7.0^{+0.1}_{-0.1} \times 10^{-3}$ & 2910 / 2548 \\
        \hline
        Absorption & & $5.9^{+0.3}_{-0.2} \times 10^{20}$ & & & & \\
        {\tt mytorus} & $1.87^{+0.00}_{-0.01}$ & $1.1^{+0.3}_{-0.2} \times 10^{23}$ & 50$^f$ & & & $7.0^{+0.0}_{-0.1} \times 10^{-3}$ & 2837 / 2529 \\
        \hline
    \end{tabular}
    \caption{Summary of continuum models. Note that these models include Gaussian lines with strengths consistent with those given in Table~\ref{tab:fe_lines}, except the {\tt mytorus} model which only has additional \ion{Fe}{xxv} and \ion{Fe}{xxvi} lines because it self-consistently includes its own neutral Fe K$\alpha$ line. The photon index and normalisation of the {\tt mytorus} component applies to the power law continuum and all components of the {\tt mytorus} model where appropriate; see the text for more details. $i$ is the inclination angle of the observer. $f$ denotes a frozen parameter.}
    \label{tab:fits}
\end{table*}

\section{Discussion}

We have considered two different scenarios to explain the overall \emph{Suzaku} and \emph{NuSTAR} observations of M81$^*$ and, in neither case, is there a detectable Compton hump, with a constraint on the reflection fraction $R < 0.11$. If the Fe~K$\alpha$ line originates in reflection off Compton-thick material, the $\sim 40$~eV EW of this line suggests that, for an illuminating continuum of $\Gamma = 1.9$ and solar abundances, the reflection fraction should be $R \simeq 0.3-0.4$ \citep[e.g.,][]{George1991}. An optically-thick disc reflection origin for the Fe~K$\alpha$ line is clearly ruled out. Furthermore, the observed limit on reflection fraction implies that the EW of the iron line should be $\lesssim 12 \text{\,eV}$, indicating that the majority of the Fe~K$\alpha$ line flux we observe does not originate from disc reflection.

We can approximate the X-ray continuum source size as $h \sim 100 r_g$ based on the variability timescale (see Section~\ref{sec:light_curve}). The weak reflection then suggests that if the disc is centrally illuminated by a source at height $h$, the inner radius $r_\text{in} \sim 20 h \sim 2 \times 10^3 r_g$ \citep{George1991}. This all suggests that the inner accretion flow within at least a $\text{few} \times 10^3 r_g$ is optically thin, and we have a very ``clean'' view of the nucleus. The X-ray continuum source size we have assumed here is much larger than $\lesssim 4 r_g$ typical of more Eddington luminous AGN \cite[e.g.,][]{Emmanoulopoulos2014}.

\begin{figure}
    \centerline{
        \begin{tikzpicture}[scale=0.55]
            % Assume a Schwarzschild black hole so horizon is at 2GM/c^2.
            \draw (-5,0) -- (-0.8,1.833);
            \draw (0.8,1.833) -- (5,0);
            \draw (5,0) -- (0.8,-1.833);
            \draw (-0.8,-1.833) -- (-5,0);
            \draw[very thick] (-5,0) -- (-6,0);
            \draw[very thick] (5,0) -- (6,0);
            \draw[very thick] (-6,0) -- (-6.866,-0.5);
            \draw[very thick] (6,0) -- (6.866,0.5);
            \draw[very thick, dotted] (-6.866,-0.5) -- (-7.732,-1.0);
            \draw[very thick, dotted] (6.866,0.5) -- (7.732,1.0);
            \draw (0,0) circle(2);
            \filldraw[fill=black] (0,0) circle(0.3);
            \node at (0,-1) {$\tau < 1$};
            \node at (-2.9,0.3) {$N_H$};
            \node at (-2.9,-0.3) {$\sim 10^{23}$};
            \node at (-5.7,-0.5) {$\tau \gg 1$};
            % Scale bar.
            \draw (0,-3) -- (7,-3);
            \node at (-2, -3.5) {$\log(r/r_g)$};
            \node at (0,-3.5) {0};
            \draw (0,-3) -- (0,-2.8);
            \node at (1,-3.5) {1};
            \draw (1,-3) -- (1,-2.8);
            \node at (2,-3.5) {2};
            \draw (2,-3) -- (2,-2.8);
            \node at (3,-3.5) {3};
            \draw (3,-3) -- (3,-2.8);
            \node at (4,-3.5) {4};
            \draw (4,-3) -- (4,-2.8);
            \node at (5,-3.5) {5};
            \draw (5,-3) -- (5,-2.8);
            \node at (6,-3.5) {6};
            \draw (6,-3) -- (6,-2.8);
            \node at (7,-3.5) {7};
            \draw (7,-3) -- (7,-2.8);
            % Observer
            \draw[dashed] (0,0) -- (-3.830,3.214);
            \draw[dashed] (0,0) -- (0,1.5);
            \draw[->] (0,1) arc (90:140:1);
            \node at (-0.5, 1.25) {$50^\circ$};
            \node at (-4.213,3.535) {Observer};
        \end{tikzpicture}
    }
    \caption{A cartoon of the possible geometry for M81$^*$. The radial scale here is logarithmic, as follows. The region $\log(r/r_g) \le 0.3$ represents a Schwarzschild black hole (although we do not know the spin of the black hole in M81$^*$). The region $0.3 < \log(r/r_g) \le 2$ represents an optically thin ($\tau < 1$), hot, inefficient flow. The region $2 < \log(r/r_g) \le 5$ represents some transition from optically thin to optically thick, the details of which are uncertain but it likely has a half-opening angle larger than $50^\circ$ and a total column density $\sim 10^{23} \text{\,cm}^{-2}$; we have modelled this component with {\sc mytorus} \citep{Murphy2009, Yaqoob2012}. This is surrounded by a warped thin disc ($\tau \gg 1$) at $\log(r/r_g) > 5$. The viewing angle of $50^\circ$ is equal to the inclination angle of the inner thin disk and the radio jet, providing a relatively unobscured view of the nucleus.}
    \label{fig:geometry}
\end{figure}
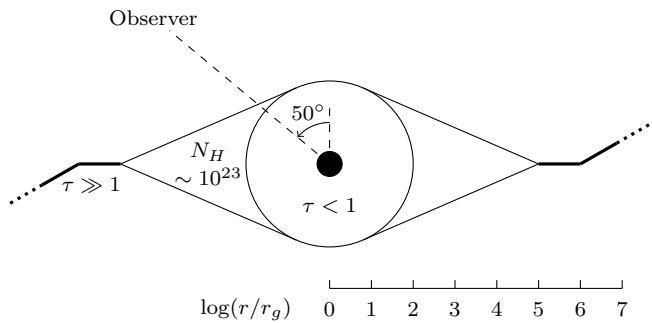

Instead of disc reflection, the Fe~K$\alpha$ line is well explained by the {\sc mytorus} model in which the continuum, scattered and iron line components originate in an optically thin ``torus''. The precise geometry of this model has not been fully explored, and is unlikely to be unique, but what we see is consistent with a torus of half-opening angle larger than the inclination angle of the inner disc, with a column density of $\sim 10^{23} \text{\,cm\,}^{-2}$. It isn't clear at what distance this material lies from the nucleus, as that is not directly constrained by the model, but the torus might correspond to the BLR. A possible geometry for M81$^*$ is shown in Fig.~\ref{fig:geometry}. The BLR in M81 is unlikely to be an outflow, as a radiatively driven outflow cannot be supported by such a weak nucleus, so a disc or spherical inflow is a more plausible geometry \citep{Devereux2007}.

\begin{figure}
    \includegraphics[width=0.45\textwidth]{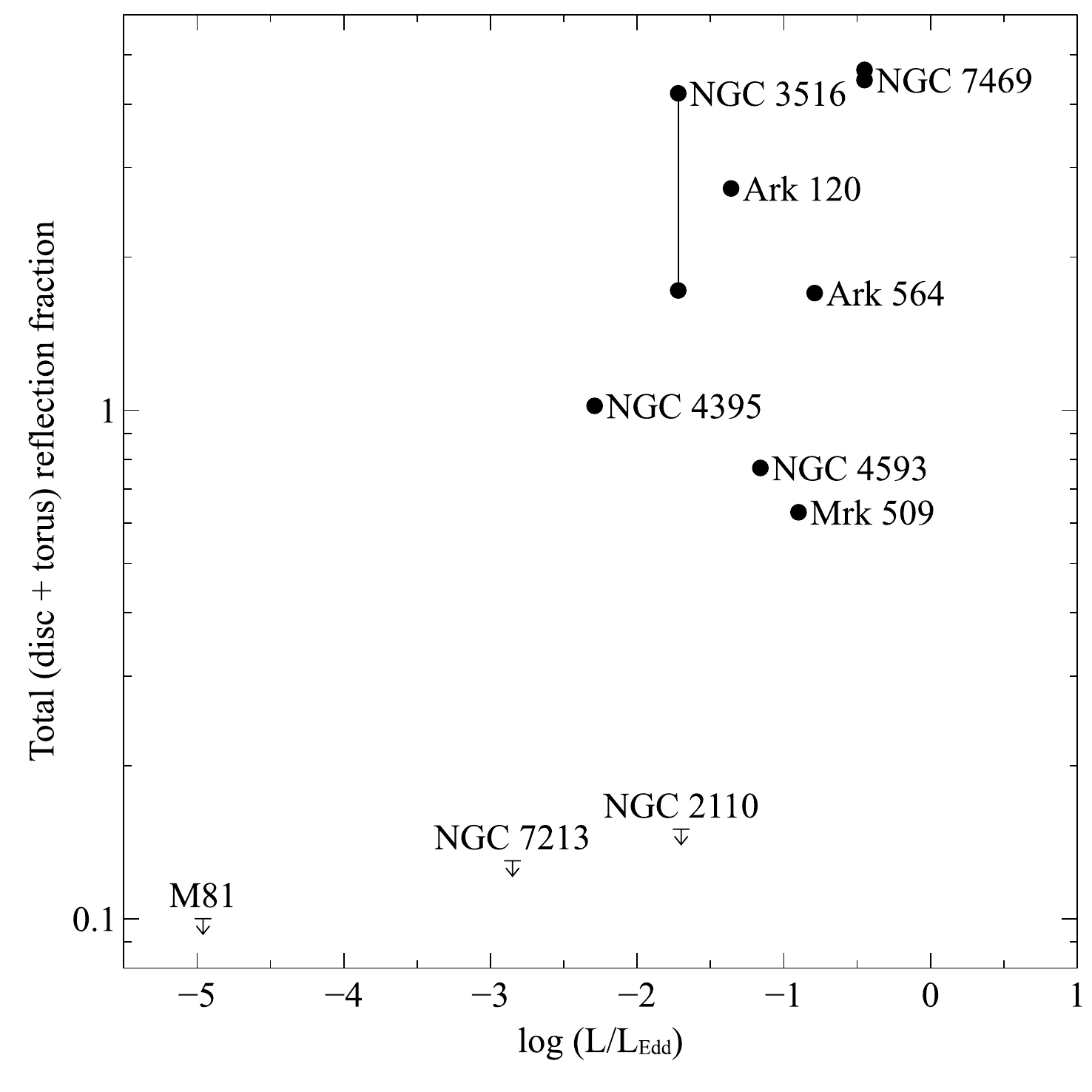}
    \caption{Reflection fraction vs. Eddington ratio. The three AGN known to unambiguously have no reprocessed emission from Compton thick material are M81 (this work), NGC 7213 
    \citep{Ursini2015}, and NGC 2110 \citep{Marinucci2014}. The other data points have reflection fractions measured with \emph{XMM-Newton} by \citet{Nandra2007a} and Eddington fractions using the bolometric luminosities and black hole masses reported by \citet{Vasudevan2007}.}
    \label{fig:ref_vs_l_edd}
\end{figure}

Our results are consistent with the hypothesis that the inner accretion environment of LLAGN is very clean, and the optically thick accretion disc is truncated at a large radius. Other similar galaxies include the Seyfert 2 galaxy NGC~2110 with $R < 0.14$ \citep{Marinucci2014} and the low accretion rate LINER NGC~7213 with $R < 0.13$ \citep{Ursini2015, Lobban2010}. These galaxies are all radiating significantly below the Eddington limit, with $L / L_\text{Edd} = 0.25 - 3.7 \times 10^{-2}$ for NGC~2110 \citep{Marinucci2014} and $L / L_\text{Edd} = 1.4 \times 10^{-3}$ for NGC~7213 \citep{Ursini2015}. What is not clear, however, is whether there are low Eddington fraction sources that have high reflection fractions. While low Eddington fraction galaxies can have fairly large EW iron lines \citep[e.g., see][]{Bianchi2008} this does not necessarily mean their reflection fractions are large -- this is not the case in NGC~2110, for example. For the AGN in the XXL survey \citep{Pierre2016} with $L / L_\text{Edd} \gtrsim 10^{-2}$ the reflection fraction is typically $R > 0.3$ \citep{Liu2016}, although since these authors have to perform a stacking analysis it isn't clear what $R$ is for the very lowest Eddington fraction galaxies. An illustration of the reflection fraction vs. Eddington fraction is shown in Fig.~\ref{fig:ref_vs_l_edd}, which suggests that the transition between high ($R > 0.3$) and low ($R \simeq 0$) reflection fraction may be quite abrupt. Higher signal-to-noise ratio observations of low Eddington fraction AGN are necessary to clarify this picture, to fill in or exclude galaxies from the upper left quadrant of this plot.

The \ion{Fe}{xxv} and \ion{Fe}{xxvi} lines may originate in hot, optically thin gas within a $\text{few} \times 10^3 r_g$. Highly ionized iron lines are also seen in NGC~7213 \citep{Lobban2010}, a galaxy with very similar spectral properties to M81$^*$. Observations of ``bare'' Seyfert galaxies also show \ion{Fe}{xxv} and \ion{Fe}{xxvi} lines \citep{Patrick2010}, suggesting that these hot iron lines may be a common component of AGN spectra. The LLAGN NGC~2110, however, does not show evidence of ionized iron lines \citep{Evans2007}, but NGC~2110 is a Seyfert~2 galaxy so does not have a ``bare'' nucleus.

\section{Conclusions}

In this paper we present simultaneous \emph{Suzaku} and \emph{NuSTAR} observations of the LLAGN M81$^*$. We find the following.

\begin{enumerate}
    \item The \emph{NuSTAR} light curve of shows significant, $\sim 30\%$, variability on timescales of $\sim 40$~ks.
    \item The continuum spectrum is well described by a cut-off power law with photon index $\Gamma = 1.9$ and cut-off energy of $E_c \simeq 220-270$~keV.
    \item There are weak Fe K$\alpha$, \ion{Fe}{xxiv} and \ion{Fe}{xxvi} lines with EWs of 40, 38 and 33~eV, respectively.
    \item There is evidence of weak intrinsic absorption, with the absorbing column density $N_H = 6.3^{+0.3}_{-0.3} \times 10^{20} \text{\,cm}^{-2}$ exceeding the Galactic value $N_H(\text{Gal}) = 4.2 \times 10^{20} \text{\,cm}^{-2}$.
    \item There is no Compton reflection hump, with the reflection fraction measured to be $R < 0.11$.
    \item We model the Fe~K$\alpha$ line as being produced in an optically thin, $N_H = 1.1 \times 10^{23} \text{\,cm}^{-2}$ torus using the {\sc mytorus} model \citep{Murphy2009, Yaqoob2012}.
    \item We hypothesise that the primary X-ray emission originates in a hot, optically thin inner accretion flow within $\sim 10^2 r_g$. Surrounding this is the ``torus'', with a half-opening angle larger than $50^\circ$, that extends from $\sim 10^2 - 10^5 r_g$. At larger radii, $\gtrsim 10^5 r_g$, is a warped, optically thick, geometrically thin disc.
\end{enumerate}

\section*{Acknowledgements}

This research has made use of the NuSTAR Data Analysis Software (NuSTARDAS) jointly developed by the ASI Science Data Center (ASDC, Italy) and the California Institute of Technology (Caltech, USA). We are grateful to Katja Pottschmidt and Kenji Hamaguchi for helpful discussion regarding the \emph{Suzaku} calibration and the problem we had with XIS~0.

\bibliographystyle{mnras}
\bibliography{mendeley}

\label{lastpage}
\end{document}